\UseRawInputEncoding
\documentclass[sigconf,screen]{acmart}
\usepackage{algorithm}
\usepackage{algorithmic}
\usepackage{booktabs}
\usepackage{amsmath}
\usepackage{amsfonts}
\usepackage{amssymb}
\usepackage{booktabs}
\usepackage{multirow}
\usepackage[normalem]{ulem}
\useunder{\uline}{\ul}{}
\AtBeginDocument{%
  \providecommand\BibTeX{{%
    \normalfont B\kern-0.5em{\scshape i\kern-0.25em b}\kern-0.8em\TeX}}}

\acmPrice{15.00}
\acmISBN{978-1-4503-XXXX-X/18/06}

\title{Hierarchical Multi-Interest Co-Network \\For Coarse-Grained Ranking}

\author{Xu Yuan, Chen Xu, Qiwei Chen, Chao Li, Junfeng Ge, Wenwu Ou}

\affiliation{%
  \institution{Alibaba Group}
  \country{Beijing, China}
 }

 \email{yunsong.yx@alibaba-inc.com}

\begin{document}

\begin{abstract}
  In this era of information explosion, a personalized recommendation system is convenient for users to get information they are interested in.
 To deal with billions of users and items, large-scale online recommendation services usually consist of three stages: candidate generation, coarse-grained ranking, and fine-grained ranking.
 The success of each stage depends on whether the model accurately captures the interests of users, which are usually hidden in users' behavior data.
 Previous research shows that users' interests are diverse, and one vector is not sufficient to capture users' different preferences. 
 Therefore, many methods use multiple vectors to encode users' interests. 
 However, there are two unsolved problems: 
 %(1) the multiple embedded vectors extracted by existing methods have high correlation and can not represent the interests of users in all aspects
 (1) The similarity of different  vectors in existing methods is too high, with too much redundant information. Consequently, the interests of users are not fully represented.
 (2) Existing methods model the long-term and short-term behaviors together, ignoring the differences between them.
 This paper proposes a Hierarchical Multi-Interest Co-Network (HCN) to capture users' diverse interests in the coarse-grained ranking stage. 
 Specifically, we design a hierarchical multi-interest extraction layer to update users' diverse interest centers iteratively.
 %The multiple embedded vectors obtained in this way are more irrelevant and can better represent users' interests in various aspects.
 The multiple embedded vectors obtained in this way contain more information and represent the interests of users better in various aspects.
 Furthermore, we develop a Co-Interest Network to integrate users' long-term and short-term interests.
 Experiments on several real-world datasets and one large-scale industrial dataset show that HCN effectively outperforms the state-of-the-art methods. 
 We deploy HCN into a large-scale real world E-commerce system and achieve extra 2.5\% improvements on GMV (Gross Merchandise Value).
\end{abstract}

\maketitle

\section{Introduction}

 Due to the severe problem of information overload in online services, recommendation systems(RS) have gradually become an important tool to reduce users' time to acquire information.
 Industrial RS usually consists of the candidate generation stage and the ranking stage. 
 The candidate generation stage aims to select items that users may be interested in from the billion corpora. 
 These items are required to be sufficiently diverse to cover all potential preferences of users.
 The ranking stage is to select the items of interest to users under a relatively small data scale and finally expose them to users.
 
\begin{figure}[!thp]
\centering
\includegraphics[scale=0.25]{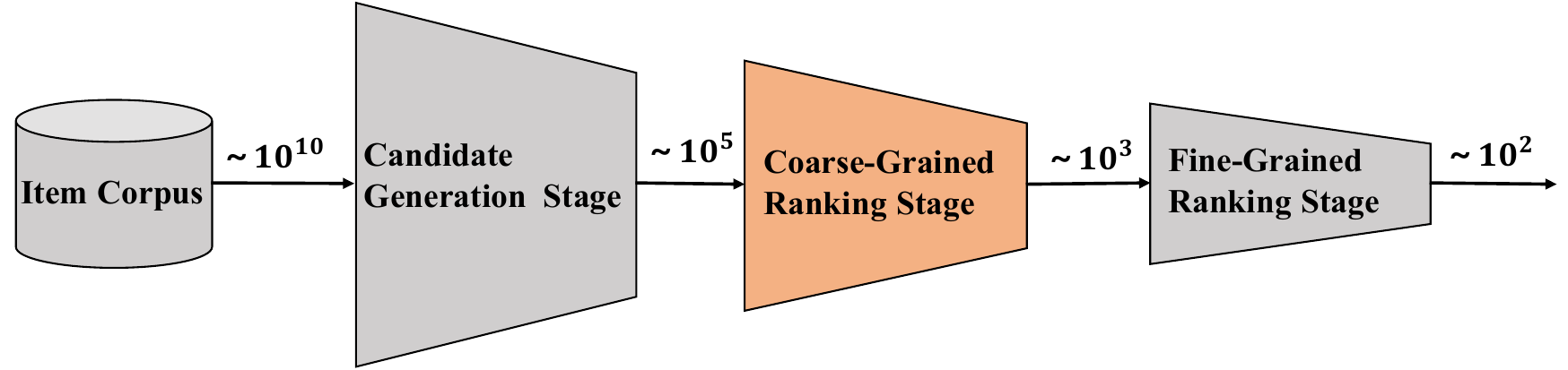}     
\caption{Overview of large industrial recommendations. The coarse-grained ranking stage is used to preliminarily score and filter the matched items, reduce data scale and pass it to the fine-grained ranking stage.}
\label{fig:rec}
\end{figure}

 For large industrial recommender systems, the ranking stage is further divided into a coarse-grained ranking stage and a fine-grained ranking stage. 
 Figure \ref{fig:rec} shows the overview.
 This multi-stage cascade architecture has been widely used, and research on coarse-grained ranking has also achieved good results\cite{Wang2020COLDTT, 10.1145/3097983.3098011}.

% Therefore, the computational complexity required in the candidate generation stage is much lower than that in the ranking stage, which is why \textit{Target Attention}(the target item is used to calculate the weight of the behavior sequence to activate the strongly correlated behaviors\cite{10.1145/3219819.3219823}) can not be used in the candidate generation stage.

Target attention mechanism proposed by DIN\cite{10.1145/3219819.3219823} is a powerful method for exploring user preferences, which has been widely used as a basic structure in the fine-grained ranking stage\cite{zhou2018deep} with hundreds of candidates.
Specifically, the candidate item is used to calculate the weight of the behavior sequence to activate the strongly correlated behaviors.
However, due to the expensive computation and storage resources, 
DIN uses the recent 50 behaviors for target attention, which ignores the user' long-term interests hidden in the long user behavior sequence and is obviously sub-optimal\cite{eta}.
Not to mention that it is not applicable to the coarse-grained ranking stage with tens of thousands of candidates.
%Not to mention that the original DIN was only applicable to the fine-grained ranking stage with only hundreds of candidates.
%Not to mention that it is applicable to the fine-grained ranking stage with only a few hundred candidate items.
%Similarly, if there are millions of candidates, each candidate item must interact with each user's behavior, which costs an unacceptable computation and storage.

Thus, in the coarse-grained ranking stage, we uses the inner product model\cite{10.1145/2505515.2505665, 48840} to avoid the delay and load caused by large-scale candidate sets.
In the inner product model, candidates no longer interact with each user's behavior, but directly calculate the matching score by inner product of the user embedding.

%YouTube DNN is a typical two tower inner product model, which is often used in the candidate generation and the coarse-grained ranking stage.

%The items representing their explicit interests in the user's historical behavior can be extracted effectively.
% However, DIN costs an unacceptable computation and storage facing the large candidate set\cite{eta}. 
% During serving, we compute all the item embeddings off-line in advance.
% When a request comes, we only need to execute one forward pass to get the user embedding and compute its inner product with all candidates, which is more efficient than DIN.
% However, , candidate items cannot be used to capture users' specific interests, due to the huge computation cost of evaluating the large candidate set.
%  In a sense, the coarse-grained ranking is a simplified version of the fine-grained ranking.
%  User-wise and item-wise vectors are pre-calculated in an offline manner with no cross features, then the inner product of the two vectors is calculated online to obtain the score. 
%  %so users’ current preferences cannot be captured with the help of target item. 

 %In the inner product model, how to capture users' interests in the above two stages is a big challenge.
%  However，In the inner product model, how to capture users' interests is a big challenge.
 However, without the help of target attention, how to capture users' interest is a huge challenge.
 A lot of methods have been proposed to capture a user's dynamic interests from his historical behaviors:
 recurrent neural networks (RNNs)\cite{hidasi2016sessionbased}, convolutional neural networks (CNNs)\cite{10.1145/3159652.3159656,10.1145/3289600.3290975}, Graph Neural Network(GNN)\cite{Wu_Tang_Zhu_Wang_Xie_Tan_2019}, self-attention mechanisms\cite{ijcai2018-546}.
 CNN handles the transition within a sliding window, whereas RNN-based methods apply GRU or LSTM to compress dynamic user interests with hidden states.
 GNN-based methods take advantage of directed graphs to model complex user-item transitions in structured relation datasets. 
 And self-attention mechanisms emphasize relevant and essential interactions with different  user-item interaction weights.
%  To further strengthen the extraction of user interests, there are also some valuable works in sequence modeling, which combines self-supervised learning\cite{ijcai2021-457, 10.1145/3340531.3411954}, aggregates side information\cite{10.1145/3404835.3463060, fdsa}, and so on.
 
%  However, these methods represent the user's interest in a single vector, and its expression ability is still limited.
 However, these methods use a single vector to represent a user's interest, leading to limited expressing ability.
 More and more studies have shown that users' historical behaviors are complex, and multiple embedded vectors express users' interests better. 
 MIND\cite{10.1145/3357384.3357814} extracts multiple interests of users with capsule network.
 SINE\cite{Tan2021SparseInterestNF} proposes a sparse-interest embedding framework which adaptively activates users' multiple intentions.
 ComiRec\cite{10.1145/3394486.3403344} uses dynamic routing and self-attention as interest extraction methods and uses an aggregation module to aggregate items from different interests.
 
 These studies have achieved good results in recommendation, yet they all have two apparent problems.
%  However, these methods are all studied in the candidate generation stage, 
%  %t present, there is no multi-interest modeling method in the coarse-grained ranking stage. 
%  This paper will be the first work of multi-interest modeling to the coarse-grained ranking stage of recommendation systems.
 %However, they all have two obvious problems.
 %Moreover, we also solve the two common problems of the above multi-interest modeling methods
 First, it is not easy to extract multiple embedded vectors from the users' behavior sequence in industry-level data. 
 Items usually do not have clear clusters. 
 %Most of the extracted multiple vectors are not independent and focused enough to truly represent the diverse interests of users. 
 The similarity and correlation of different vectors extracted by existing methods are too high, resulting in a lot of redundant information. 
 Therefore, the performance of these multi-vector modeling methods is not much better than that of the single-vector modeling methods, because the extracted vectors can not really represent the diverse interests of users\cite{8594844,Tan2021SparseInterestNF}.
%  Although the multi-layer transformer can extract higher-level features from the behavior sequence, it is found in the experiment that such transformation is not helpful to capture the user's interests better than the recommended single header implementation\cite{8594844,Tan2021SparseInterestNF}. re
 %Second, when people start shopping on mobile TaoBao, their behavior will accumulate into a relatively long sequence. 
 Second, when people start surfing the Internet, their behavior will accumulate into a relatively long sequence. 
 This long-term behavior expresses users' potential preferences. 
 At the same time, the more recent behaviors better represent the users' precise  interests\cite{10.1145/3109859.3109896}. 
 For example, if a user is a comic fan, he may click on a large number of comic-related items before. 
 When he chooses to buy clothes now, clothes co-branded with comics will attract her/him more than other clothes. 
 Therefore, it is crucial to consider both long-term behaviors and short-term interests.
 All existing multi-interest modeling methods directly model the sequence, and ignore this internal structure.
 Some studies\cite{ijcai2019-585,hgn,sdm} have also noticed the importance of the cross features of users' long-term and short-term interests. But they are all based on single-vector modeling. On the aspect of extracting user preferences with multiple vectors, the existing methods are not effective.
%  Cartesian product can capture the cross characteristics of long-term and short-term interests\cite{zhou2020can}, but the memory and computational overhead are unacceptable.
 
 In order to overcome the limitations of existing methods, we propose the Hierarchical Multi-Interest Co-Network(HCN) to capture users' diverse interests in the coarse-grained ranking stage.  
 Firstly, the user's long-term and short-term historical behaviors are encoded respectively.
 And then they are continuously updated through hierarchical multi-layer extraction modules. 
 %Through iterative updating, the captured interest centers can be more accurate and focused.
 Through iterative updating, the captured interest centers is more accurate and diverse, and their redundant information will be gradually reduced.
 Secondly, a Co-Interest Network is constructed to integrate users' long-term and short-term preferences, which captures the cross features. 
 %We let short-term intentions activate the corresponding information in long-term behaviors and let long-term intentions activate short-term behaviors to capture cross features.
 Finally, we build an aggregation module to aggregate multiple interest centers obtained by Hierarchical Multi-Interest Co-Network.
 To summarize, the main contributions of this work are as follows:
 
 \begin{itemize}
\item We propose a hierarchical multi-interest aggregation network, which encodes more accurate and diverse interests of users.

\item We consider both short-term and long-term behaviors. These two parts are modeled separately, which represent different levels of user interests. A Co-Interest module effectively combines long-term preferences and current interests, which incorporates their correlation information rather than simple combinations.

% \item Our HCN model is evaluated on five offline datasets in the real world and outperforms the other state-of-the-art methods. As the first model applied in coarse-
% grained ranking stage in the industry, we successfully deployed it on production environment of recommender system at Taobao. The HCN model achieves significant improvements compared to previous online system.
% \end{itemize}

% \item Experiments on four large-scale datasets and a real industrial dataset show that our model HCN outperforms several state-of-the-art baselines. As the first model applied in coarse-grained ranking stage in the industry, we successfully deployed it on production environment of recommender system at Taobao. The HCN model achieves significant improvements compared to previous online system.
\item Experiments on four large-scale datasets and a real industrial dataset show that our model HCN outperforms several state-of-the-art baselines. We successfully deployed it on a real world E-commerce system in September 2021 and achieved achieve extra 2.5\% improvements on GMV (Gross Merchandise Value). 
%The HCN model achieves significant improvements compared to previous online system.
\end{itemize}

\section{Methodology}

% \subsection{Problem Statement}

%  Suppose $\mathcal{I}$ is the set of items and $\mathcal{U}$ is the set of users. For a specific user $u\in\mathcal{U}$, list $I_u=[q_1^{(u)},q_2^{(u)}...,q_k^{(u)}]$ denote the corresponding interaction sequence in chronological order where $q_k\in\mathcal{I}$ and $k$ is the time step. 
%  Given the interaction history $I_u$, sequential recommendation aims to predict the item that user $u$ will interact with at time step $k + 1$. 

%\subsection{Coarse-Grained Ranking at TaoBao}
\subsection{Coarse-Grained Ranking for Industrial}
%  To better understand the Hierarchical Multi-Interest Co-Network in this work, we firstly give an overview of TaoBao recommendations in Figure \ref{fig:rec}.
 %To better understand the Hierarchical Multi-Interest Co-Network in this work, we firstly give an overview of large E-commerce recommendations.
 For sizeable industrial recommendation systems like E-commerce, goods and users are often billions of levels. 
 %For example, mobile Taobao has a daily life of 300 million. 
 In such a large amount of data, coarse-grained ranking as a filter is critical in modern RS.

\begin{figure}[!thp]
    \centering
    \includegraphics[scale=0.2]{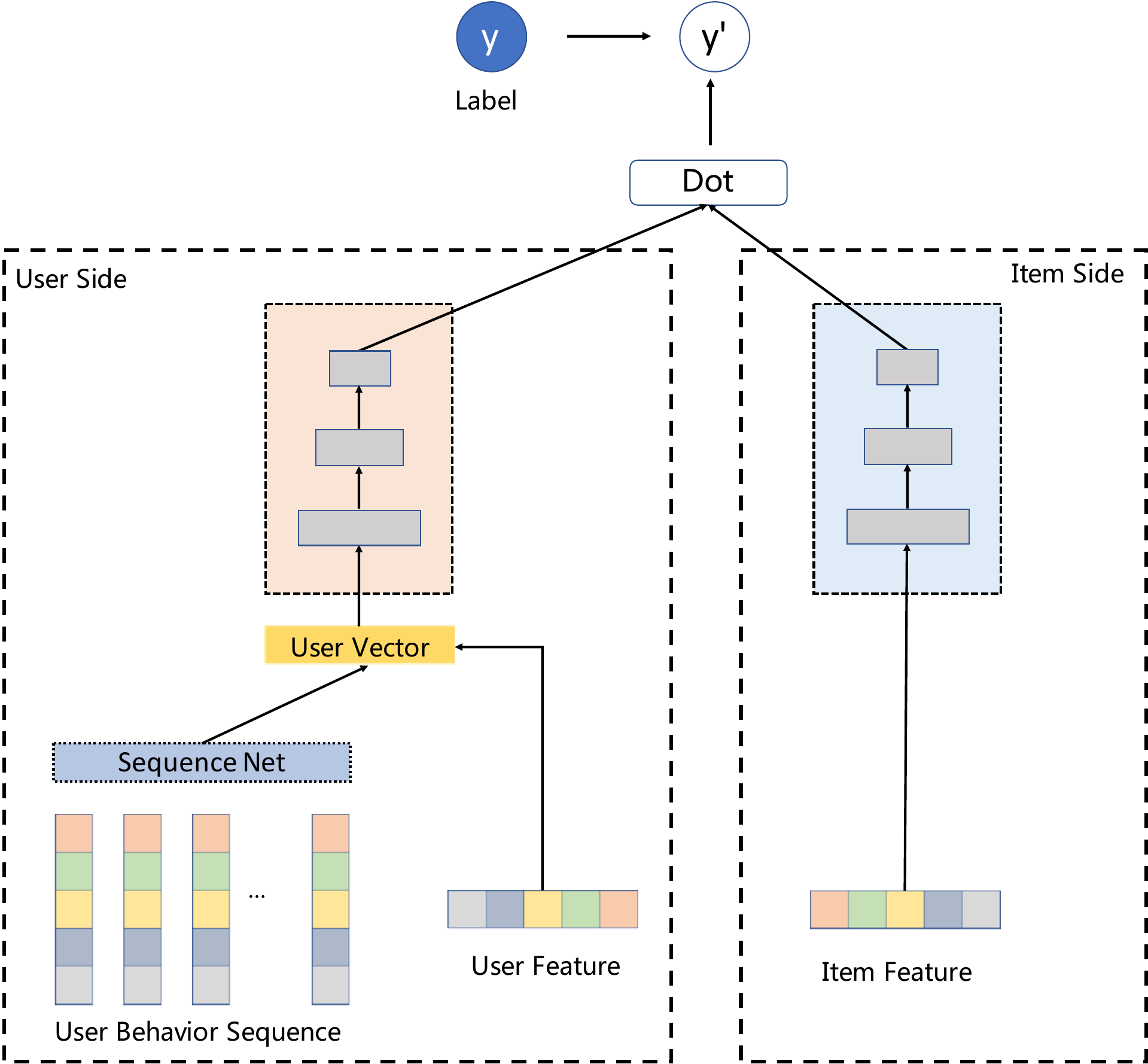}     
    \caption{Industrial Coarse-Grained Ranking Architecture.}
    \label{fig:double}
\end{figure}

 %copy from pfd
 In the coarse-grained ranking stage, we mainly estimate the CTRs of all items selected in the candidate generation stage\cite{10.1145/3394486.3403309}, and then use these CTRs scores to choose the top-$k$ items for the next stage. The inputs of the prediction model are mainly composed of three parts. The first part includes user behavior, which records the history of items clicked by users. The second part includes user features, such as user ID, age, gender, etc. The third part includes item features, such as item ID, category, etc. 
 %In this work, all features are transformed into categorical types, and we learned an embedding of each one.
%  In the coarse-grained ranking stage, we are mainly to estimate the CTRs of all items selected by the candidate generation stage, which are then used to select the top-k highest ranked items for the next stage. The inputs of the prediction model mainly consist of three parts. The first part consists of the user behavior, which records the history of her clicked/purchased items. The second part consists of the user features, e.g., user id, age, gender, etc. and the third part consists of the item features, e.g., item id, category, brand, etc. Across this work, all features are transformed into categorical type and we learn an embedding for each one.
 %  In the coarse-grained ranking stage, the complexity of the prediction model is strictly restricted, in order to grade tens of thousands of candidates in milliseconds. Here we utilize the inner product model\cite{10.1145/2505515.2505665} to measure the item scores:
Due to the large scale of the candidate set, the complexity of the prediction model and computing resources are strictly limited. Here, we use the inner product model\cite{10.1145/2505515.2505665,48840} to measure the item scores:
 \begin{equation}
 \label{eqn:double}
 f(\mathbf{X}^U, \mathbf{X}^I) \triangleq \langle \Phi_{W^U}(\mathbf{X}^U), \Phi_{W^I}(\mathbf{X}^I)\rangle,
 \end{equation}
where the superscript $U$ and $I$ denote the user and item, respectively. $X^{U}$ means a combination of user behavior and user features. $\Phi_{W}(\cdot)$ represents the non-linear mapping with learned parameter $W$. 
$\langle\cdot, \cdot\rangle$ is the inner product operation. 
As the user side and the item side are separated in Equation (\ref{eqn:double}), during serving, we can compute the mappings $\Phi_{W^I}(\cdot)$ of all items off-line in advance. 
When a request comes, we only need to execute one forward pass to get the user mapping $\Phi_{W^U}\left(\mathbf{X}^U\right)$ and compute its inner product with all candidates, which is highly efficient. 
For more details, see the illustration in Figure \ref{fig:double}. 
We conclude the notations and the corresponding descriptions in Table \ref{Table:notation}.
 %copy from pfd

 %Considering the balance between efficiency and accuracy, the double tower structure referred to DSSM\cite{10.1145/2505515.2505665} is adopted.
 %In this structure, target item cannot interact with the user interaction sequence at the bottom like DIN\cite{10.1145/3219819.3219823} as they greatly increase the latency at serving, but can only be matched by inner product at the top.
 %Figure \ref{fig:double} is the architecture diagram of coarse-grained ranking stage. 
 
%  The output of the user tower is obtained by stitching the statistical characteristics of the user and the characteristics obtained from the modeling of the user's historical behavior through a three-layer neural network, and the output of the item tower is obtained by the statistical characteristics of the item through a three-layer neural network. 
%  In the upper layer, the matching degree is calculated by inner product operation.

 \begin{figure*}[!thp]
    \centering
    \includegraphics[scale=0.18]{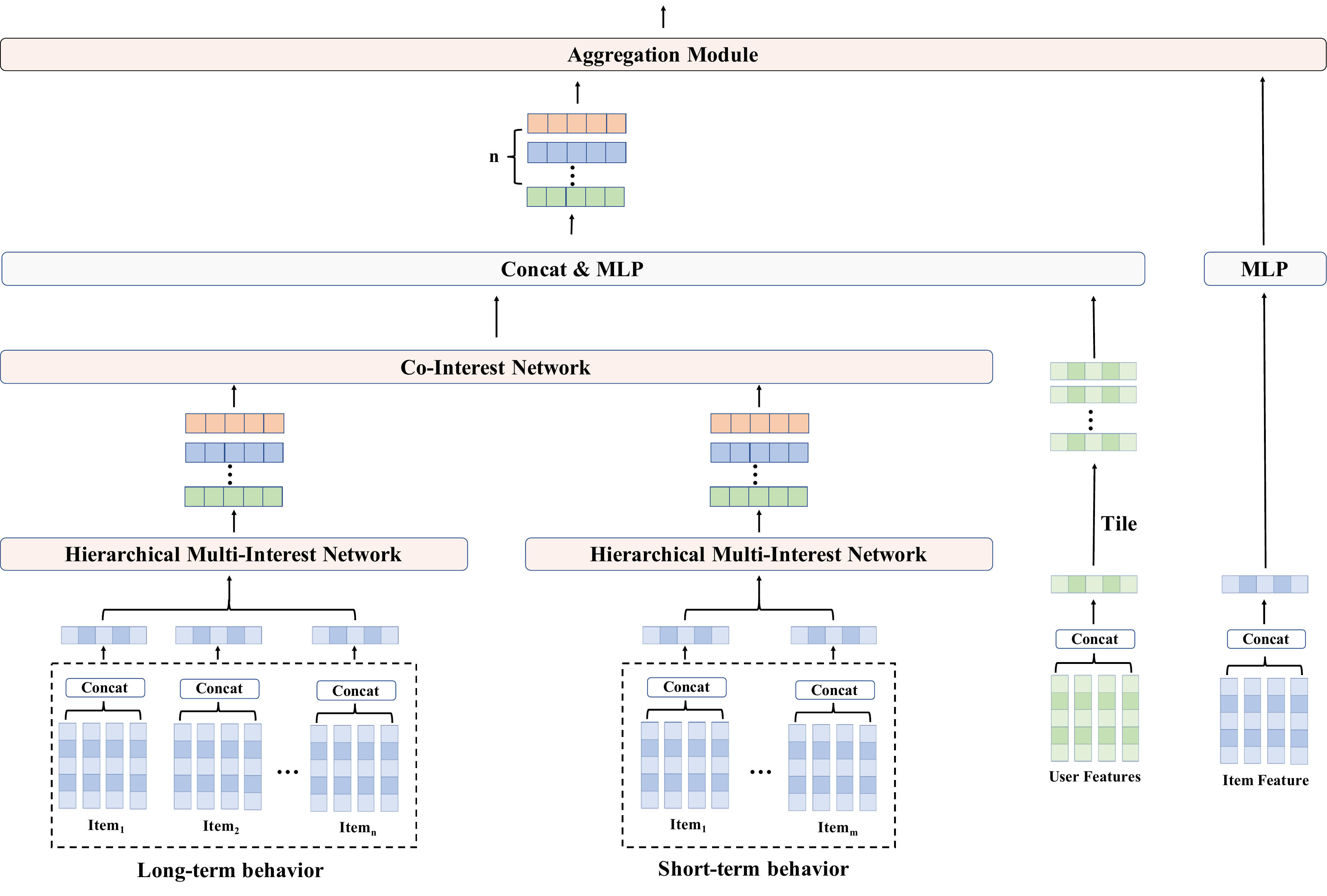}     
    \caption{Hierarchical Multi-Interest Framework.}
    \label{fig:bestmodel}
\end{figure*}

\subsection{Hierarchical Multi-Interest Framework}

 The success of personalized recommendations depends on whether the model accurately describes the users’ interests, which are usually hidden in the users’ historical behaviors.
 Some studies show that the bottleneck of accurately capturing users' interests is to express them through one representation vector\cite{10.1145/3357384.3357814}.
 Therefore, many methods use attention\cite{Tan2021SparseInterestNF, 10.1145/3394486.3403344}, transformer\cite{8594844}, capsule network\cite{10.1145/3357384.3357814}, etc. to capture or activate a variety of different interests of users. 
%  However, these methods may limit the ability of the model to capture multiple intentions, because the way they extract determines that the encoded interest centers are loose and unrepresentative.
 
 %In TaoBao, each user interacts with hundreds of products every day. 
 In large E-commerce, each user interacts with hundreds of products every month\cite{sim,ren2019lifelong}. 
 Interactive products often belong to different categories, which indicates the diversity of users' interests.
%  Therefore, the diverse interest centers of users can not be completely gathered through a simple attention layer.
 Therefore, we propose HCN, which updates the embedding vectors iteratively to obtain more users' interests.
 
 Specifically, We use two Hierarchical Multi-Interest Modules to extract interests from users' long-term and short-term interaction sequences, respectively. 
 After extracting multiple interests from each sequence, they are fused through a Co-Interest Network and then concated with user embedding to represent the user side. 
 In the upper layer, we design an aggregation module to integrate users' multiple interests and match them with item embedding.
 Figure \ref{fig:bestmodel} illustrates the Hierarchical Multi-Interest Framework.

% \usepackage{booktabs}
\begin{table}[!ht]
\caption{The notations and the corresponding descriptions.}
\centering
\resizebox{1\columnwidth}{!}{
\begin{tabular}{@{}cl@{}}
\toprule
Notation & Description  \\ \midrule
\ $n$ & The number of user interest centers.  \\
\ D & The dimension of the embeddings. \\
\ $\hat{\mathbf{L}}$  & The embedding of users' long-term interests. \\
\ $\hat{\mathbf{S}}$ & The embedding of users' short-term interests.  \\
\ \multirow{2}{*}{$\mathbf{X}^U$} & The output on user side, a combination of user \\
                                  & behavior and user features.\\
\ $\mathbf{X}^I$ & The output on item side. \\ 
\ \multirow{2}{*}{$\tilde{\mathbf{X}}^U$} & The embedded vector encoded by user \\
                                          & history behavior sequence.\\
\ \multirow{2}{*}{$\hat{\mathbf{X}}^U$} & The embedded vector formed by user statistical \\
                                        &features.\\
\ $\ell_{S}$  & The length of user long-term behavior sequence. \\
\ $\ell_{L}$ & The length of user short-term behavior sequence.  \\ 
\ $C_0$  & Input user interest Centers. \\
\ $F_c$  & Output user interest Centers. \\
$\Phi_{W}(\cdot)$ & The non-linear mapping with learned parameter $W$\\
\bottomrule
\end{tabular}}

\label{Table:notation}
\end{table}

\subsubsection{Hierarchical Multi-Interest Network}
 Our framework starts by extracting users' diverse interests from the users' historical behavior sequences. 
 %We build a hierarchical multi-interest network.
 The basic structure is Multi-Head Attention, which has strong feature extraction ability and has been verified by many studies\cite{clark2019does,vig2019multiscale}.
 However, it is found in the experiment that such transformation is not helpful to capture the users' interests better than the recommended single header implementation\cite{8594844,Tan2021SparseInterestNF}.

%  Therefore, unlike SASRec\cite{8594844}, we do not use the multi-head mechanism to represent multiple interests but just use it to extract the features of the sequence.

Poly-Encoders\cite{Humeau2020PolyencodersAA} proposed a new architecture with an additional learnt attention mechanism that represents more global features from sequence. 
We use it to extract multiple vectors from historical behavior sequence as the user's interests, which are represented with $n$ vectors$(y_{user}^1...y_{user}^n)$ instead of just one.
To obtain these $n$ global representations, we learn $n$ context codes$(c_1...c_n)$ which are randomly initialized.
Specifically, we obtain $y_{user}^i$ using:

\begin{equation}
\label{eqn:poly}
y_{user}^{i}=\sum_{j} \omega_{j}^{c_{i}} q_{j}
\end{equation}
where $\left(\omega_{1}^{c_{i}}, . ., \omega_{N}^{c_{i}}\right)=\operatorname{softmax}\left(c_{i} \cdot q_{1}, . ., c_{i} \cdot q_{N}\right)$, $q_{j}$ represents each token of the sequence, $N$ is the length of the sequence.

% The 

% initialize $k$ vectors as global features.
% To obtain these $k$ global features that represent the input, we learn m context codes (c1, ..., cm), where ci extracts representation y
% i
% ctxt
% by attending over all the outputs of the previous layer. 

% 具体的，本文随机初始化了$k$个向量作为$Query$向量，与$Key$向量做内积运算求得每个Token的权重$\omega$，然后对$Value$向量做加权求和。在这里，$Value$向量即为用户历史交互序列，Token则代表用户交互的每一个行为，本质上是一个个项目的嵌入式表示。最后得到的$k$个向量$[y_{user}^1, ..., y_{user}^k]$即为编码得到的用户历史行为序列的表示。其中$y_{user}^i$的计算方式如公式 (\ref{eqn:poly})所示，$h_i$代表用户历史交互的每一个项目的嵌入式表示，$c_i$代表每个行为的权重。

 For the long-term and short-term behavior sequences, we initialize vectors $(c_1...c_n)$ as matrix $\mathbf{C_0}\in \mathbb{R}^{n\times D}$,which called the \textbf{SEED MATRIX}. We take the user history interaction sequence as matrix $\mathbf{K} \in \mathbb{R}^{\ell_{L}\times D}$ and matrix $\mathbf{V} \in \mathbb{R}^{\ell_{L}\times D}$(for short sequence, they are matrix $\mathbf{K} \in \mathbb{R}^{\ell_{S}\times D}$ and matrix $\mathbf{V} \in \mathbb{R}^{\ell_{S}\times D}$) to encode interests respectively. 
 %for Multi-Head Attention calculation respectively. 
 The calculation formula can be expressed as Equation (\ref{eqn:mha}):
 \begin{equation}
 \label{eqn:mha}
 \begin{aligned}
 H &= Concat(head_1,...,head_h)W^o, \\
 head_i &= softmax(\frac{\mathbf{C_0}W_i^C(\mathbf{K}W_i^K)^T}{\sqrt{d_{\mathbf{K}W_i^K}}})\mathbf{V}W_i^V, 
 \end{aligned}
 \end{equation}
 where $H$ represents the encoded user interest centers, $W_i^C$,$W_i^K$,$W_i^V$ and $W^o$ are learnable parameters.

 \renewcommand{\algorithmicrequire}{\textbf{Input:}}
\renewcommand{\algorithmicensure}{\textbf{Output:}}
\begin{algorithm}[htb] 
    %\setstretch{1.35}
    \caption{Hierarchical Multi-Interest calculation process.}
    \label{alg:Framwork} 
    \begin{algorithmic}[1]
        \REQUIRE 
            User history behavior sequence $I \gets [q_1,q_2...,q_n]$;  
            Iteration times $r$;  
            head num $h$;  
            Input user interest Centers $C_0$;  
        \ENSURE
            Output user interest Centers $F_c$;  
        % \STATE {$\text{Random Initialize } \left( c_1...c_k \right) \test{\quad as \quad} C_0$}
        \STATE {$\text{Random Initialize } C_0$}
        \FOR{$i=0 \to r$}
            \IF{$i>0$}
                \STATE {$C_i \gets H_{i-1}$}
            \ENDIF
            \FOR{$k=0 \to h$}
                \STATE {$\text{Init } W_k^Q, W_k^K, W_k^V$}
                \STATE {$Q \gets C_iW_k^Q$}
                \STATE {$K \gets IW_k^K$}
                \STATE {$V \gets IW_k^V$}
                \STATE {$head_k \gets softmax(\frac{QK^T}{\sqrt{d_{K}}})V$}
            \ENDFOR
            \STATE {$H_i \gets Concat(head_1,...,head_h)W^o$}
        \ENDFOR
        \STATE $F_c \gets W_c[H_0;H_1;...H_r]+b_c $
        \RETURN{$F_c$}
    \end{algorithmic}
\end{algorithm}
 
 Compared with the traditional self-attention, the calculation method we designed can reduce the complexity from $\ell_{L}^2$ to $ n\ell_{L}$ where $n \ll \ell_{L}$. 
 And it is also acceptable for ultra-long sequences. 
 
 %However, such a simple calculation is not enough to obtain compact diverse users' interest centers. 
 However, such a simple calculation is not enough to obtain diverse and representative users' interest centers.
 Researches on image representation learning show that the features become sharper and more distinguishable with the deepening of the network\cite{10.1145/3065386,zeiler2014visualizing}.
 Similar observations are also found in sentence representation learning\cite{raganato-tiedemann-2018-analysis}.
 %The initialization of interest centers will bring great uncertainty to the result.
%  The interest extraction modules of many methods only stay at this step, which leads to the loose and unrepresentative interests extracted by them.
 Inspired by previous studies, we further introduce hierarchical structure as the extraction layer to get more representative interest centers.

 We regard multi-interest extraction based on multi-head attention as an iteration.
 After one iteration, the interest centers are global, which can not sufficiently represent the differentiation of users' interests. 
 Therefore, we use a hierarchical structure to update the users' interest centers in multiple rounds. 
 Starting from the second layer, we use the interest centers output from the previous layer as $\mathbf{Q}$ and continue to encode the users' historical behavior sequence as Equation (\ref{eqn:mha}).
 Finally, the output of each layer is concated together, and the users' interest centers are obtained after a linear transformation.
 Through such iterative calculation, we get more diverse and representative interest centers.
 The whole hierarchical calculation process is listed in Algorithm \ref{alg:Framwork}.

\begin{figure}[!ht]
    \centering
    \includegraphics[scale=0.18]{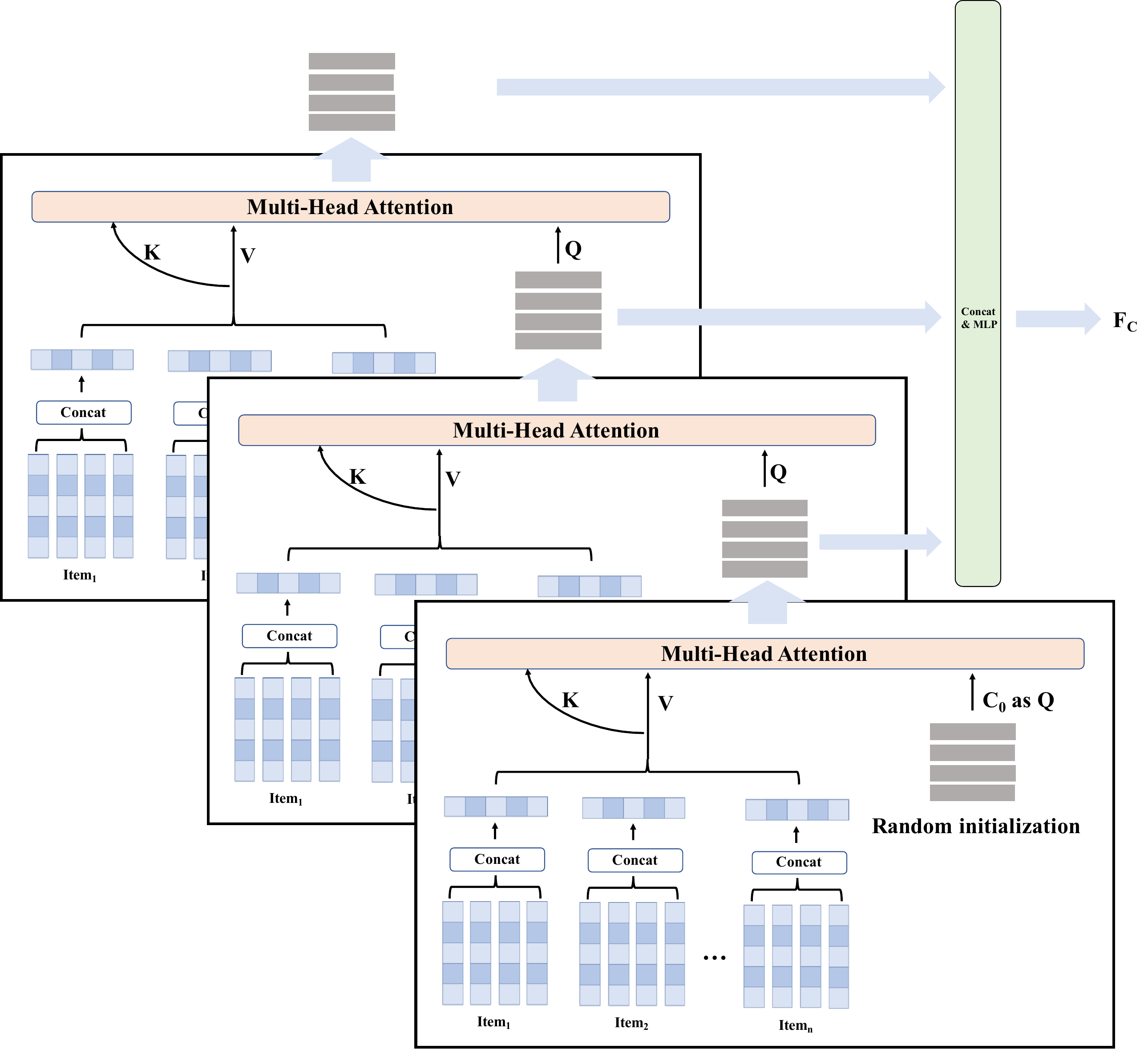}     
    \caption{Hierarchical Multi-Interest Network.}
    \label{fig:dieceng}
\end{figure}

 Through iterative multi-layer calculation, the final aggregated multiple vectors can represent the different interests of users.
 We get two vectors $\hat{\mathbf{L}} \in \mathbb{R}^{n\times D}$ and $\hat{\mathbf{S}} \in \mathbb{R}^{n\times D}$ to represent users' long-term interests and short-term interests respectively.
 Figure \ref{fig:dieceng} illustrates the Hierarchical Multi-Interest Network.

\subsubsection{Co-Interest Network}
 Till now, we have extracted users' long-term and short-term interests.
 Next, we need to integrate them.
 In the single-vector modeling method, we use a simple concat operation. 
 However, since two sets of multi-vectors are extracted, how to align them has become an important problem.
 DIN\cite{10.1145/3219819.3219823} use a novel designed local activation unit to help behaviors with higher relevance to the candidate item get higher activated weights and dominate the representation of user interests. 
 However, their approach is only applicable to the fine-grained ranking stage, which is not appropriate to the coarse-grained ranking stage with a larger sample size and strict complexity constraints.
 The cartesian product can capture the cross features of long-term and short-term interests\cite{zhou2020can}, but the memory and computational overhead are also unacceptable.

 In this paper, we design a Co-Interest Network to integrate the long-term and short-term interests with low consumption of computing resources. 
 We regard the long-term interests as the users' potential preference and the short-term interests as the users' current preference. 
 The goal is to capture the relationship between the potential and the current intentions.
 %Specifically, we argue for levering attention instead of Cartesian product to capture the interactive information between long-term and short-term interests. 
%  Figure \ref{fig:coin} illustrates the Co-Interest Network.

%  \begin{figure}[!thp]
%     \centering
%     \includegraphics[scale=0.19]{cin.pdf}     
%     \caption{Co-Interest Network.}
%     \label{fig:coin}
% \end{figure}

 Firstly, the local inference is determined by the attention weight $e_{ij}$ computed as Equation (\ref{eqn:weight}), which is used to obtain the local relevance between long-term and short-term interests.
%  Firstly, the short-term interests are regarded as Q, and the long-term interests are regarded as K and V for Multi-Head Attention calculation. 
%  The corresponding information is filtered and strengthened from the long-term interests through current strong interests signals. 
%  Then, the long-term interests are regarded as Q, and the short-term interests are regarded as K and V for Multi-Head Attention calculation. 
%  The corresponding information is filtered and strengthened from the short-term interests through stable and continuous interest signals.
 \begin{equation}
 \label{eqn:weight}
 e_{ij} = \hat{\mathbf{L}}_i\hat{\mathbf{S}}_j^T .
 \end{equation}
 
 Then we use $e_{ij}$ to reconstruct short-term and long-term interests, respectively. 
 The motivation is to give higher weight to the embedding vector related to short-term intention and capture the cross features of long-term and short-term interests.
 %is to find the information of current interests from potential preferences, and find the corresponding potential interests from current preferences, and then let them guide each other and learn more fully. 
 The reconstruction process is as shown in Equation (\ref{eqn:ls}):
 \begin{equation}
 \label{eqn:ls}
 \begin{aligned}
 \tilde{\mathbf{L}}_{i} &=\sum_{j=1}^{\ell_{S}} \frac{\exp \left(e_{i j}\right)}{\sum_{k=1}^{\ell_{S}} \exp \left(e_{i k}\right)} \hat{\mathbf{S}}_{j}, \forall i \in\left[1, \ldots, \ell_{L}\right], \\
\tilde{\mathbf{S}}_{j} &=\sum_{i=1}^{\ell_{L}} \frac{\exp \left(e_{i j}\right)}{\sum_{k=1}^{\ell_{L}} \exp \left(e_{k j}\right)} \hat{\mathbf{L}}_{i}, \forall j \in\left[1, \ldots, \ell_{S}\right].
 \end{aligned}
 \end{equation}

 Finally, the fusion vectors are concated together and output, which defined as:
 \begin{equation}
 \label{eqn:ouput}
 \begin{aligned}
\tilde{\mathbf{X}}^U=W^h[\hat{\mathbf{L}} ; \tilde{\mathbf{L}} ; \hat{\mathbf{S}} ; \tilde{\mathbf{S}}]+b^h.
 \end{aligned}
 \end{equation}
 Here, the embedded vector $\tilde{\mathbf{X}}^U \in \mathbb{R}^{n\times D}$ represent $n$ users' interest centers which contains the cross features of users' long-term and short-term preference. 
%  Compared with the target attention proposed by DIN\cite{10.1145/3219819.3219823}, our proposed method reduces the complexity from linear to constant for each user's recommendation.
 The embedded vector formed by users' statistical features $\hat{\mathbf{X}}^U\in \mathbb{R}^{1\times D}$, so we tile it into $n$ parts and concat it with $\mathbf{X}^U$ as the output of user side, which defined as:
  \begin{equation}
 \label{eqn:ouputend}
 \begin{aligned}
 \mathbf{X}^U = W^U[\tilde{\mathbf{X}}^U;\hat{\mathbf{X}}^U]+b^U.
 \end{aligned}
 \end{equation}

\subsubsection{Aggregation Module}
 After obtaining the user side representation, we need to match it with the item side representation.
 %Under the framework of the coarse grained ranking, target item cannot interact with the underlying user behavior sequence like DIN\cite{10.1145/3219819.3219823}. 
 We build an aggregation module layer to allow the candidate item $X^I$ to interact with multiple representations $X^U$ gotten on the user side at the top level, so as to make full use of the information contained in the candidate item. 
 The calculation process is shown in Equation (\ref{eqn:top}):
 \begin{equation}
 \label{eqn:top}
 \begin{aligned}
 \alpha_j &= \frac{\exp((\mathbf{X}^I)^T\mathbf{X}^U)}{\sum_{j'=0}^n\exp((\mathbf{X}^I)^T\mathbf{X}^U_{j'}))}, \\
 y &= \langle \mathbf{X}^I,  \sum_{j=0}^k\alpha_{j} \mathbf{X}^U_{j}\rangle,
 \end{aligned}
 \end{equation}
 where $\langle\cdot, \cdot\rangle$ is the inner product operation.
 The model is trained by minimizing the loss of cross-entropy, which is defined as:
  \begin{equation}
 \label{eqn:loss}
 \begin{aligned}
 \mathbf{\mathcal{L}} = -\hat{y}log(\frac{1}{1+e^{-y}}) - (1 - \hat{y})log(1 - \frac{1}{1+e^{-y}}),
 \end{aligned}
 \end{equation}
 where $\hat{y}$ is label, and $y$ is the output of the model, that is, the matching score.

% \subsubsection{Online Serving}
%  For online serving, we use our multi-interest extraction module to compute multiple interests for each user. Each interest vector of a user can independently retrieve top-N items from the large-scale item pool by the nearest neighbor library such as Faiss [25]. The items retrieved by multiple interests are fed into an aggregation module to determine the overall item candidates. Finally, the items with higher ranking scores will be recommended for user.

\section{Experiments}
% In this section, we conduct experiments at four benchmark datasets and one industrial dataset with the aim of answering the following research questions:
% %and one billion-scale industrial data in Alibaba with the aim of answering the following research questions:
% % over three benchmark datasets and one billion-scale industrial data to validate the proposed approach. 
% \begin{itemize}
%  \item \textbf{RQ1:} What is the performance of HCN compared to other state-of-the-art baselines?
%  %\item \textbf{RQ2:} What information can HCN extract from multiple vectors?
%  \item \textbf{RQ2:} What are the effects of the different modules, Hierarchical Multi-Interest Network, and Co-Interest Network through ablation studies?
%  %\item \textbf{RQ3:} What is the difference between user' diverse interests obtained by HCN and those obtained by other state-of-the-art methods?
%  \item \textbf{RQ3:} Is HCN sensitive to the hype-parameter $n$?
%  \item \textbf{RQ4:} Can HCN be used in the candidate generation stage?
%  \end{itemize}

\subsection{Experimental Settings}
\subsubsection{Data Sets}
 We conduct our experiments on five real-world datasets from several global E-commerce or other platforms. 
 The data statistics after pre-processing are listed in Table \ref{Table:data}.
 We truncated and padded the long-term and short-term behavior sequences for different datasets.
 These pre-processing method has been widely used in related works\cite{10.1145/3340531.3412744,10.1145/3292500.3330984}.
 
 \begin{itemize}
 \item \textbf{Amazon Dataset} consists of product reviews and metadata from Amazon\footnote{http://jmcauley.ucsd.edu/data/amazon/}. In our experiment, we use the Books category of the Amazon dataset. And we select the data from October 2010 to October 2018.
  \item \textbf{MovieLens 25m}\cite{10.1145/2827872} is a large stable benchmark which consists of movie ratings\footnote{https://grouplens.org/datasets/movielens/25m/}. The original data contains 25 million ratings and one million tag applications applied to 62,000 movies by 162,000 users.
 \item \textbf{Taobao} is a dataset consisting of user behavior data retrieved from Taobao\footnote{https://tianchi.aliyun.com/dataset/dataDetail?dataId=649}, one of the biggest e-commerce platforms in China. It contains user behaviors from November 25 to December 3, 2017, with several behavior types, including click, purchase, add to cart, and item favoring. 
 \item \textbf{Tafeng Dataset} released on Kaggle, which contains the transaction data of Chinese grocery store\footnote{https://www.kaggle.com/chiranjivdas09/ta-feng-grocery-dataset}. It contains user behaviors from November 1, 2000 to February 28, 2001. 
 %collects user behaviors from Taobao’s recommender systems\footnote{https://tianchi.aliyun.com/dataset/dataDetail?dataId=649}. In our experiment, we only use the click behaviors    
%  \item \textbf{Industrial Dataset} is collected by mobile Taobao, one of the biggest e-commerce platforms in China. We chose the data from April 24, 2021 to May 22, 2021 as the training data. The total scale of the data is about 1600TB, and the ratio of positive and negative samples is 1:20. The user's long-term click sequence length is 500 and the short-term click sequence length is 50.
 \item \textbf{Industrial Dataset} is collected by one of the world's largest e-commerce platforms. We chose the data from April 24, 2021 to May 22, 2021 as the training data. The total scale of the data is about 1600TB, and the ratio of positive and negative samples is 1:20. 
\end{itemize}

% \usepackage{booktabs}
\begin{table}[!ht]
\caption{The statistics of five datasets after pre-processing.}
\centering
\resizebox{1\columnwidth}{!}{
\begin{tabular}{@{}lccccc@{}}
\toprule
Dataset                & Taobao       & Amazon Books  &   ML25m    &  Tafeng & Industrial \\ \midrule
\# Users               &  140,240     &  14,561       &   11,686    &  8,154  & 278 million \\
\# Items               &  1,080,770   & 535,181       &   31,287   &  22,437 &  37 million\\
\# Actions             & 32,120,591   &  5,107,694    &   2,517,391  &   475,790 &  14,000 million \\
\# Avg. Actions / User &      229     &   351         &   215  &   58   & 50  \\
\# Avg. Actions / Item &      30      &   10          &   80  &   21  & 378  \\ 
\# Long-term sequence length  & 120    & 120   & 80    & 20     & 500        \\
\# Short-term sequence length & 30     & 30    & 20    & 5      & 50         \\ 
\bottomrule
\end{tabular}}

\label{Table:data}
\end{table}

\subsubsection{Evaluation Metric}
We evaluate the performance with the area under the ROC curve (AUC).
% AUC\cite{FAWCETT2006861} is used for the evaluation of binary classification and is not affected by the distribution of sample categories. 
AUC\cite{FAWCETT2006861} is a standard metric in coarse-grained ranking in industry\cite{10.1145/3397271.3401440,10.1145/3437963.3441824,10.1145/3394486.3403309}, which measures the goodness of order by ranking all items with predicted CTR.
%The higher the AUC, the better the performance of the model at distinguishing between the positive and negative classes.

% \usepackage{booktabs}
\begin{table*}[h!t]
\caption{Performance (AUC) comparison of baselines and our approaches, where our approach HCN's best results are in bold. The underlined numbers are the best results besides HCN.}  % 表格标题
\centering
%\resizebox{1\columnwidth}{!}{
\begin{tabular}{@{}lcccccc@{}}
\toprule
                                     & Taobao & Amazon Books &   ML25m    &  Tafeng & Industrial \\ \midrule
Mean Pooling                         & 0.7320 &    0.6989    &   0.8248   &  0.7505       &     0.7218      \\
GRU4Rec                              & 0.7544 &    0.7167    &   0.8304   &  0.7630         &   0.7382   \\
SAMRec                               & 0.7691 &    0.7189    &   0.8340   &  0.7619        &        0.7407   \\
SASRec                               & 0.7613 &    0.7205    &   0.8337   &  0.7611        &   0.7395   \\
MIND                                 & 0.7706 &    \underline{0.7293}    &   0.8357   &  0.7690        &    0.7447     \\
SINE                                 & 0.7632 &    0.7284    &   0.8343   &  \underline{0.7730}       &      0.7445  \\
ComiRec                              & \underline{0.7730} &    0.7267    &   \underline{0.8358}   &  0.7724        &     \underline{0.7451}    \\
%Multi-Interest Network               & 0.7682 &    0.6642         &            &         &   0.7449  \\
HCN                                  & \textbf{0.7767} &    \textbf{0.7341} &  \textbf{0.8385}  &    \textbf{0.7785}&    \textbf{0.7482}   \\ \midrule
Imprv.                               & 3.7\textperthousand      &      4.8\textperthousand        &     2.7\textperthousand       &   5.5\textperthousand      &  3.1\textperthousand   \\ \bottomrule
\end{tabular}
%}

\label{table2}  % 用于索引表格的标签
\end{table*}

\subsubsection{Baselines}
\begin{itemize}
\item \textbf{Mean Pooling} is a simple online baseline for aggregating users' historical behaviors, which has the advantages of low complexity and low computation.

\item \textbf{GRU4Rec}\cite{hidasi2016sessionbased} adopts GRU to capture sequential dependencies and makes predictions for session-based recommendation.

\item \textbf{SASRec}\cite{8594844} uses stacking self-attention to capture user interests in user historical interaction behavior.

\item \textbf{SAMRec}\cite{attention} is another common online baseline to extract users' interests. Different from SASRec, it first obtains the high-level semantics of the sequence through the stacked self-attention and then aggregates them through mean pooling.

\item \textbf{MIND}\cite{10.1145/3357384.3357814} models the user historical behavior sequences with capsule network and uses dynamic routing to aggregate them into the interest expression vector adaptively. It shows promising results in the candidate generation stage.

\item \textbf{SINE}\cite{Tan2021SparseInterestNF} is a novel sparse interest embedding framework, which can adaptively activate multiple intentions for each user.

\item \textbf{ComiRec}\cite{10.1145/3394486.3403344} uses dynamic routing and self-attentive method as extraction methods, with an aggregation module to aggregate items from different interests.
% \item \textbf{Multi-Interest Network}\cite{} is the latest multi-head attention based multi-interest extraction model in Taobao.

\end{itemize}

%   % Please add the following required packages to your document preamble:
% % \usepackage{booktabs}
% \begin{table}[!ht]
% \centering
% \resizebox{0.5\columnwidth}{!}{
% \begin{tabular}{@{}lcccc@{}}
% \toprule
%  & long-term& short-term\\ \midrule
% Taobao              &  120 &  30    \\
% Amazon Books       &  120 &  30 \\
% ML25m             & 80  & 20  \\
% Tafeng &      20     &   5       \\
% Industrial &      500     &   50       \\ 
% \bottomrule
% \end{tabular}}
% \caption{The statistics of selected sequence length.}
% \label{Table:len}
% \end{table}

% \begin{table}[!ht]
% \centering
% \resizebox{1\columnwidth}{!}{
% \begin{tabular}{cccccc}
% \toprule
%                           & Taobao & Books & ML25m & Tafeng & Industrial \\ \midrule
% long-term sequence length  & 120    & 120   & 80    & 20     & 500        \\
% short-term sequence length & 30     & 30    & 20    & 5      & 50         \\ \bottomrule
% \end{tabular}}
% \caption{The statistics of selected sequence length.}
% \label{Table:len}
% \end{table}

\subsubsection{Parameter Settings}
 For Mean Pooling and SAMRec, we implement them with PyTorch. 
 For other methods, we use the source code provided by their authors. 
 All hyper-parameters are set following the suggestions from the original papers. 
 For HCN, we set the number of attention heads as 4.
 %in public datasets and 8 in the industrial dataset.
 The number of users' interests is 8 in the public datasets and 16 in the industrial dataset.
 In the hierarchical multi-interest extraction module, considering the trade-off between efficiency and accuracy, we finally stack four layers.
 The dimension of the embedding is 48 in the public datasets and 128 in the industrial dataset.
 %We use the Adam optimizer\cite{adam} with a learning rate of 0.001.

\subsection{Results and Analysis}

 \subsubsection{Offline Results}
 Table \ref{table2} presents the performance comparisons between several baselines and our model (HCN). 
 For fairness, the number of embedded vectors obtained by all multi-vector modeling methods is the same.
 HCN achieves the best performance on four public datasets and one industrial dataset, verifying our model's superiority.
 HCN performs better than Mean Pooling, GRU4Rec, and SAMRec because HCN can capture a variety of users' interests.
 It can also be observed that employing multiple embedding vectors (MIND, ComiRec, SINE, HCN) for a user generally performs better than single-embedding-based methods (Mean Pooling, GRU4Rec, SAMRec, SASRec).
 
 Compared with MIND, our model shows much better performance. 
 On the one hand, the infrastructures of the two methods are different. 
 We use Multi-Head Attention as the basis, which is better than the dynamic routing used by MIND.
 In complex behavior sequences, the feature extraction ability of the Multi-Head Attention mechanism plays a very important role.
 It is also verified in ComiRec\cite{10.1145/3394486.3403344}.
 On the other hand, in our model, long-term and short-term interests interact and guide each other.
 While MIND does not model long-term and short-term preference, respectively, thus ignoring the cross features.
%  One reason is that the infrastructure is different. We use Multi-Head Attention as the base, which is better than the dynamic routing method of MIND. 
%  This is also verified in ComiRec\cite{10.1145/3394486.3403344}.
%  In our model, long term and short-term interests interact and guide each other. 
%  MIND does not model long-term and short-term interests separately. 
%  And MIND has achieved good results in the matching stage, but it is not suitable in the coarse-grained ranking stage with smaller data scale. 
 
 HCN outperforms SINE and ComiRec. 
 We conjecture that the hierarchical structure leads to more diverse and representative of multiple embedded vectors extracted by the model.
 To verify it, we introduce the Pearson correlation coefficient, which reflects a linear correlation of variables.
 It is the ratio between the covariance of two variables and the product of their standard deviations.
 The lower the Pearson correlation coefficient, the lower the correlation between multiple embedding vectors.
%  When the correlation coefficient is 0, the obtained multiple vectors are orthogonal. 
 The Pearson correlation coefficient calculation process is as shown in Equation (\ref{eqn:r}):
 \begin{equation}
 \label{eqn:r}
 r(X, Y)=\frac{\operatorname{Cov}(X, Y)}{\sqrt{\operatorname{Var}[X] \operatorname{Var}[Y]}}
 \end{equation}
 where $\operatorname{Cov}(X, Y)$ is the covariance between $X$ and $Y$, $\operatorname{Var}[X]$is the variance of $X$, and $\operatorname{Var}[Y]$ is the variance of $Y$.
 
 \begin{table}[!h]
 \caption{Correlation Coefficient comparison between various state-of-the-arts models and HCN. The lower the value, the lower the correlation between embedded vectors.}  %
\centering
\resizebox{1\columnwidth}{!}{
\begin{tabular}{@{}lcccccc@{}}
\toprule
                                     & Taobao & Amazon Books &   ML25m    &  Tafeng &  Industrial\\ \midrule
SINE                                 & 0.8859 & 0.7023       & 0.8051     & 0.8586   &  0.7994       \\
ComiRec                              & 0.6697 & 0.6696       & 0.8136     & 0.8421    &  0.8769     \\
MIND                                 & 0.5963 & 0.6645       & 0.7702     & 0.8084  &  0.5587\\ 
HCN                                  & 0.5485 & 0.6269       & 0.7523     & 0.7965  &  0.5019\\ 
 \bottomrule
\end{tabular}
}
\label{table4}  % 用于索引表格的标签
\end{table}

\begin{table*}[!ht]
\caption{Effect of Hierarchical Multi-Interest Network (HIN) and Co-Interest Network (CIN).}  % 表格标题
\centering
\resizebox{2\columnwidth}{!}{
\begin{tabular}{ccccccccccc}
\hline
& \multicolumn{2}{c}{Taobao}  & \multicolumn{2}{c}{Amazon Books} & \multicolumn{2}{c}{ML25m}   & \multicolumn{2}{c}{Tafeng} & \multicolumn{2}{c}{Industrial} \\
                              & AUC       & Improve   & AUC      & Improve    & AUC       & Improve   & AUC     & Improve     & AUC     & Improve   \\ \hline
w/o HIN and CIN               & 0.7711    & -         &  0.7254  & -          &  0.8344   & -         & 0.7705  & -           & 0.7449  & -         \\
w/o CIN                       & 0.7742    & 3.1\textperthousand      &  0.7293  &   3.9\textperthousand     &  0.8359   &  1.5\textperthousand     & 0.7772  &  6.7\textperthousand       & 0.7472  & 2.3\textperthousand      \\
HCN                           & \textbf{0.7767} & \textbf{5.6\textperthousand} & \textbf{0.7341}    & \textbf{8.7\textperthousand}   & \textbf{0.8385} & \textbf{4.1\textperthousand} & \textbf{0.7785}  &   \textbf{8.0\textperthousand}  & \textbf{0.7482}&\textbf{3.3\textperthousand}      \\ \hline
\end{tabular}}

\label{table3}  % 用于索引表格的标签
\end{table*}

 We calculate the multiple interest vectors obtained by the model in pairs and get the overall correlation coefficient score after averaging on the whole testing set.
 Table \ref{table4} presents the result.
 It can be seen that the multiple user interest centers captured by HCN are the most independent. 
 MIND as another method of iteratively updating the user interests, the correlation of embedded vectors is also lower than SINE and ComiRec.
 %It is proved that the hierarchical structure captures sharper features.
 %Therefore, comparing SINE and ComiRec, HCN uses hierarchical structure to iteratively update the users' interest centers, and finally get the users' more focused interests.
 %Therefore, HCN has a better performance in recommendation than SINE and ComiRec. 
 The slight correlation between vectors means that there are few redundant features, which can contain more information.
 It is consistent with the goal of multi-vector modeling.
 Combined with the experimental results, we find that the embedded vector with lower correlation can often provide more incredible help for the recommendation.
 It is the fundamental reason why HCN is superior to SINE and ComiRec.
%  In order to make the comparison of correlation coefficients more intuitive, we draw a heat map of user coupling coefficients to compare the quality of user interests obtained by different methods in Figure \ref{fig:hot}. 
%  It can be seen that iterative updating methods such as MIND and HCN can obtain more representative embedded vectors

 \subsubsection{Online A/B Testing}
%  We deploy HCN on Mobile Taobao App. 
%  Compared with the previous generation main model, HCN has an overall average improvements of 3\%, 3\%, 3\%, 2.5\% in IPV, CTR, Order Number, and GMV.
%  Note that our multi-vector model has been working well online since October 2021.
 
 %As the first model applied in coarse-grained ranking stage in the industry, 
 We have deployed the proposed solution in mobile e-commerce platforms. From 2021-08-21 to 2021-09-21, we conduct a strict online A/B testing experiment to validate the proposed HCN model. 
 Compared to our last online model based on SAMRec, HCN achieves great gain, which shows in Table \ref{Table:6}. 
 Now, HCN has been deployed online and serves the Homepage of Taobao app every day, which contributes significant business revenue growth.
% Now, HCN has been deployed online and serves and contributes significant business revenue growth.

% \usepackage{booktabs}
\begin{table}[!h]
\caption{HCN’s Lift rate of online results compared with
previous online system from Aug 21 to Sep 21, 2021, in Guess
What You Like column of Taobao App Homepage.}
\centering
\resizebox{0.6\columnwidth}{!}{
\begin{tabular}{@{}cccccc@{}}
\toprule
& Metric    & IPV   & CTR   & GMV & \\ \midrule
& Lift rate & 3.2\% & 3.1\% & 2.5\% & \\ \bottomrule
\end{tabular}
}

\label{Table:6}
\end{table}

 \subsubsection{Case Study}
\begin{figure}[!ht]
    \centering
    \includegraphics[scale=0.5]{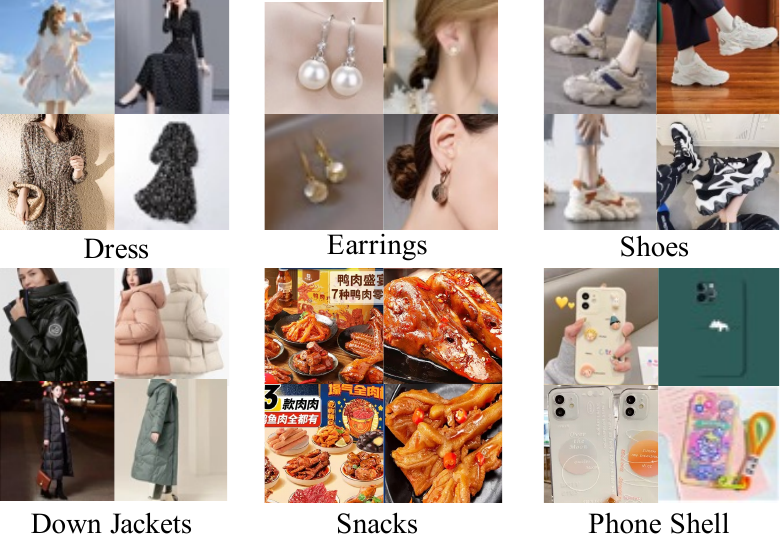}     
    \caption{A case study of an e-commerce user. We draw six interest centers, "Dress", "Earrings", "Shoes", "Down Jackets", "Snacks", "Phone Shell" with the top-4 closest items.}
    \label{fig:show}
\end{figure}

 In order to deeply explore the knowledge learned by the model, we visualized multiple interest centers.
 Concretely, we leverage its embedding vectors to retrieve the top-4 closest items under their similarity for each center. 
 Figure \ref{fig:show} illustrates six exemplar centers to show their clustering performance. 
 As can be seen, our model successfully groups some semantic-similar items into a latent concept. 
 Multiple embedded vectors can capture various commodity sets from the user's historical behaviors and then give different weights through candidate items. 
 %In this way, users' interests are dynamically activated, improving the recommendation performance and reducing computational complexity. 
 %It is very effective in the coarse-grained ranking and candidate generation stage.
 
%   \begin{figure}[!ht]
%     \centering
%     \includegraphics[scale=0.6]{hot.pdf}     
%     \caption{A case study of an e-commerce user. We draw six interest centers "Dress", "Earrings", "Shoes", "Down Jackets", "Snacks", "Phone Shell" with the top-4 closest items.}
%     \label{fig:hot}
% \end{figure}

 \subsubsection{Ablation Study}

 We perform the ablation study to see the effectiveness of the proposed Hierarchical Multi-Interest Network (HIN) and Co-Interest Network (CIN). 
 Table \ref{table3} reports the results on different datasets. 
 It can be seen that both HIN and CIN significantly improve the recommendation performance.
 HIN continuously updates users' interest centers through iteration aggregation to make them more diverse and representative. 
%  This model is superior to the existing multi-interest extraction models. 
 CIN focuses on the potential connection between long-term and short-term preferences ignored by the existing models, which significantly strengthens the expression ability of $n$ embedded vectors.
 It emphasizes the strong correlation between a small number of long-term preferences and current short-term behaviors, while captures multi-level user preferences.
 In particular, even without the help of the Co-Interest Network, our proposed model can obtain better results than state-of-the-art.

%  \subsubsection{Comparative Experiment(RQ3)}
%  % internal evaluation
%  In order to evaluate the quality of the multi embedding vectors extracted by each model, we introduce Correlation Coefficient, which are often used to evaluate the degree of correlation between variables.
%  The lower the correlation coefficient, the lower the correlation between user interest vectors.
%  When the correlation coefficient is 0, the obtained multiple vectors are orthogonal. In a sense, we just want to get such embedded vectors because they are the most representative
%  The correlation coefficient calculation process is as shown in Equation (\ref{eqn:ls}):
%  \begin{equation}
%  \label{eqn:r}
%  r(X, Y)=\frac{\operatorname{Cov}(X, Y)}{\sqrt{\operatorname{Var}[X] \operatorname{Var}[Y]}}
%  \end{equation}
%  where $\operatorname{Cov}(X, Y)$ is the covariance between $X$ and $y$, $\operatorname{Var}[x]$is the variance of $X$, and $\operatorname{Var}[Y]$ is the variance of $Y$.
 
%  We calculate the multiple interest vectors obtained by the model in pairs, and get the overall correlation coefficient score after averaging.
%  It can be seen that the multiple user interest centers captured by HCN are the most independent. 
%  We conjecture that the reason why HCN gets the best overall performance is closely related to the low correlation of getting the user interest center.

%  Table \ref{table4} presents the evaluation of multiple embedded vectors obtained by each state of the arts model. 
%  HCN has absolute advantages in each metric, which proves that the model proposed in this paper is very effective.

   \begin{figure}[!h]
    \centering
    \includegraphics[scale=0.3]{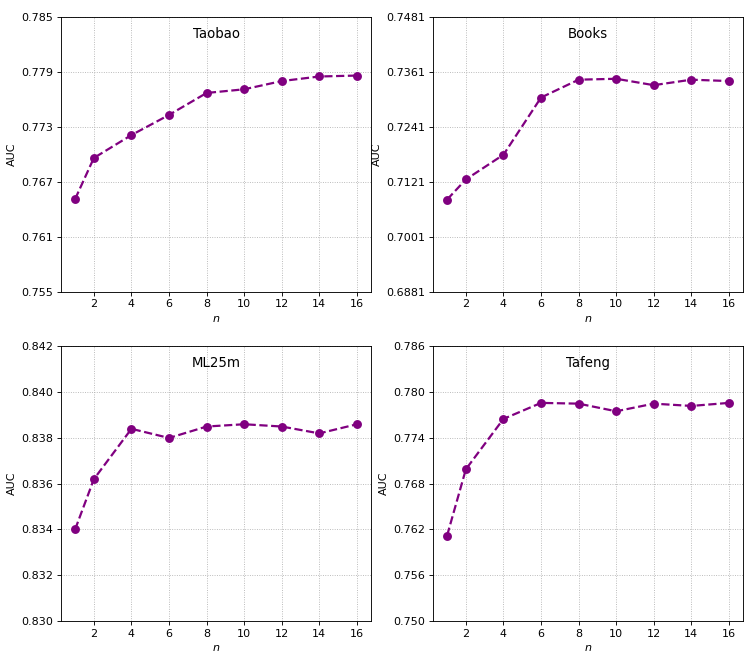}     
    \caption{Sensitivity of Hyper-parameter $n$.}
    \label{fig:n}
\end{figure}
 
 \subsubsection{Sensitivity of Hyper-parameter}
 We study the sensitivity $n$ of the HCN hyper-parameter. 
 %We want to explore the optimal number of embedded vectors that can represent user interest in different datasets.
 Figure~\ref{fig:n} shows AUC scores of HCN with different $n$.
 It can be seen that when the number of embedded vectors increases from one to multiple, AUC increases significantly, which once again proves the importance of multi-vector modeling.
 %Therefore, it is very necessary to extract multiple vectors of user interest in the coarse-grained sorting stage.
 In particular, different datasets have different sensitivity to the hyper-parameter $n$. 
 Large E-commerce datasets such as Taobao and Amazons Books need eight or more embedded vectors to represent users' interests because of their rich user historical behaviors.
 However, when $n$ continues to rise, AUC does not increase obviously. 
 The user's interests have an upper limit, which can be expressed by a specific number of embedded vectors, and more vectors are redundant.
\begin{table*}[!h]
\caption{Performance on two public datasets, where our approach HCN's best results are in bold. The underlined numbers are the best results besides HCN. All the numbers in
the table are percentage numbers with `\%' omitted.}
\centering
\resizebox{1.2\columnwidth}{!}{
\begin{tabular}{@{}lcccccccc@{}}
\toprule
                     & \multicolumn{4}{c}{Amazon Books}                                 & \multicolumn{4}{c}{Taobao}                                        \\
\multicolumn{1}{l}{} & \multicolumn{2}{c}{Metrics@50} & \multicolumn{2}{c}{Metrics@100} & \multicolumn{2}{c}{Metrics@50}  & \multicolumn{2}{c}{Metrics@100} \\ \midrule
                     & Hit Rate       & NDCG          & Hit Rate       & NDCG           & Hit Rate       & NDCG           & Hit Rate       & NDCG           \\ \midrule
Mean Pooling         & 1.54           & 0.35          & 2.31           & 0.42           & 8.73           & 3.14           & 10.03          & 3.18           \\
GRU4Rec              & 1.70           & 0.51          & 2.74           & 0.67           & 9.41           & 3.60           & 12.43          & 4.08           \\
SAMRec               & 3.07           & 0.98          & 4.22           & 1.17           & 13.55          & 5.71           & 16.19          & 6.94           \\
SASRec               & 3.17           & 1.01          & 4.43           & 1.28           & 13.36          & 5.64           & 15.37          & 6.38           \\
MIND                 & 3.94           & 1.45          & 5.87           & 1.66           & {\ul 17.81}    & {\ul 10.31}    & {\ul 20.55}    & {\ul 10.93}    \\
SINE                 & {\ul 4.36}     & {\ul 1.59}    & {\ul 6.14}     & {\ul 1.72}     & 16.37          & 9.28           & 18.96          & 9.52           \\
ComiRec              & 4.01           & 1.52          & 5.98           & 1.67           & 16.45          & 9.56           & 19.18          & 10.06          \\
HCN                  & \textbf{4.76}  & \textbf{1.64} & \textbf{6.52}  & \textbf{1.80}  & \textbf{18.21} & \textbf{11.09} & \textbf{21.96} & \textbf{11.95} \\ \bottomrule
\end{tabular}
}

\label{Table:5}
\end{table*}

 \subsubsection{Results of HCN in candidate generation stage}
 Although HCN is developed based on the coarse-grained ranking stage, it can also adapt to the candidate generation stage. 
 We employ Hit Rate and NDCG \cite{ndcg} to evaluate the recommendation performance.
 Hit Rate indicates the proportion of cases when the rated item is amongst the top-k items.
 NDCG is the normalized discounted cumulative gain at k, which takes the rank of recommended items into account and assigns larger weights on higher positions.
%  To avoid high computation cost on all user-items in evaluation, following the strategy in \cite{ijcai2019-190, Yuan2021ImprovingSR}, we randomly draw 99 negative items that have not been engaged with the user and rank the ground-truth item among the 100 items.
 To evaluate the recommendation's performance, we split each dataset into training/validation/testing sets. 
 We hold out the last two interactions as validation and test sets for each user, while the other interactions are used for training.
 
 Due to the space limit, only the results on Amazon Books and Taobao are reported in Table~\ref{Table:5}. 
 %he results are shown in Table \ref{Table:5}.
 The comparison results show that HCN consistently performs better on two datasets in terms of all evaluation metrics than state-of-the-art models.
 Similar results are obtained for other cases.
 It is consistent with our experimental results on the coarse-grained ranking stage.
 HCN, MIND, SINE, and ComiRec perform better than the single embedded modeling methods represented by GRU4Rec in Hit Rate and NDCG. 
 This further shows that it is necessary to use multiple embedded vectors to express users' preferences in a system with highly complex user behaviors.

\section{Related Work}

\subsection{Sequential Recommendation}

%copy from SSI
 Nowadays, many approaches have been proposed to model the user's historical interaction sequence. 
 The methods based on the Markov chain predict the subsequent user interaction by estimating the probability of transfer matrix between items\cite{10.5555/647235.720264}.
 RNN-based methods model the sequential dependencies over the given interactions from left to right and make recommendations based on this hidden representation. 
 Except for the basic RNN, long short-term memory (LSTM)\cite{10.1145/3018661.3018689}, gated recurrent unit (GRU)\cite{hidasi2016sessionbased}, hierarchical RNN\cite{10.1145/3109859.3109896} have also been developed to capture the long-term or more complex dependencies in a sequence.
 CNN-based methods first embed this historical interactive information into a matrix and then use CNN to treat the matrix as an image to learn its local features for subsequent recommendation\cite{10.1145/3159652.3159656,10.1145/3289600.3290975}. 
 GNN-based methods first build a directed graph on the interaction sequence, then learn the embeddings of users or items on the graph to get more complex relations over the whole graph \cite{Wu_Tang_Zhu_Wang_Xie_Tan_2019}. 
 Attention models emphasize those important interactions in a sequence while downplaying those that have nothing to do with user interest\cite{ijcai2018-546}.
 To further enhance the expression ability of sequence model, there are also some valuable works in sequence modeling, which combines self-supervised learning\cite{ijcai2021-457, 10.1145/3340531.3411954}, aggregates side information\cite{10.1145/3404835.3463060, fdsa}, and so on.
%  To further enhance the expression ability of sequence model, there are also some valuable works in sequence modeling, which combines self-supervised learning\cite{ijcai2021-457, 10.1145/3340531.3411954}, aggregates side information\cite{10.1145/3404835.3463060}, and so on.
%  In this paper, we model the user historical behavior sequence as embedded vectors and regard it as a new feature on the user side so as to improve the performance of recommendations.

\subsection{Attention Mechanism}
 Attention was first applied in the image field\cite{28419,10.5555/3045118.3045336,10.5555/2969033.2969073} for dynamic control in identifying objects, and then widely used in seqs2seqs tasks, such as machine translation\cite{Bahdanau2015NeuralMT, luong-etal-2015-effective}. 
 Traditional methods, such as RNN and CNN, can not reflect the importance of different sequence parts, and the attention mechanism can capture it well.
 In natural language processing, BERT\cite{bert} has achieved great success by using transformer. 
 In recommendation, it also plays a significant role. 
 For example, DIN\cite{10.1145/3219819.3219823} uses attention to obtain information strongly related to the candidate item in the user's historical behavior sequence. 
 ICAI-SR\cite{10.1145/3404835.3463060} uses the attention mechanism to aggregate various attributes of items.
%  In our work, we use the powerful feature extraction ability of the Multi-Head attention mechanism and reduce its computational complexity. 
%  A hierarchical structure is further designed to extract users' interests.

\subsection{Multi Interest Modeling}
 In previous studies, we expressed the user's interest as a single embedded vector.
 With the development of the Internet and the accumulation of data, more and more studies show that users' behavior history is dynamic, and users' interests are diverse.
 It isn't easy to completely represent the genuine interest of users with a single vector.
 
% Nowadays, more and more studies show that it isn't easy to completely represent the genuine interest of users with a single vector.
 MIND\cite{10.1145/3357384.3357814} uses dynamic routing to aggregate the user's historical behavior sequence into the user's interest expression.
 It does not capture the cross features of long-term and short-term preferences.
 SINE\cite{Tan2021SparseInterestNF} proposes a sparse-interest embedding framework which can adaptively activate users' multiple intentions.
 ComiRec\cite{10.1145/3394486.3403344} uses dynamic routing and self-attentive as interest extraction methods and uses an aggregation module to aggregate items from different interests.
%  The multiple interests extracted by these two methods are loose and unrepresentative.
 The difference of the proposed HCN from existing Multi-Interest models is in that we built the extraction module hierarchically and noticed the relationship between long-term and short-term interests. 
 These make the performance of the recommendation system get a significant improvement.

\section{Conclusion}

 In this work, we extract multiple interests from users' behavior sequences with Hierarchical Multi-Interest Co-Network for the coarse-grained ranking stage.
 First, users' long-term and short-term interests are extracted by a Hierarchical Multi-Interest Network and updated iteratively to get more diverse and represent centers.
 Then a Co-Interest Network captures the cross features of users' long-term and short-term preferences and integrates them through an aggregation module.
 Experimental results and analysis demonstrate the superiority of the proposed model.
 The total GMV are improved by 2.5\% in online A/B test compared to previous online system. 
 %HCN has been deployed online to serve the mainstream traffic.

% \begin{acks}
% \end{acks}

\bibliographystyle{ACM-Reference-Format}
\bibliography{references}

% \appendix


\end{document}